\documentclass[preprint,aps,superscriptaddress]{revtex4}
\usepackage{graphicx}
\usepackage{amsmath}
\usepackage[usenames]{color}
\usepackage{amssymb}
\newcommand{\be}{\begin{equation}}
\newcommand{\ee}{\end{equation}}
\newcommand{\bea}{\begin{eqnarray}}
\newcommand{\eea}{\end{eqnarray}}
\newcommand{\pr}{\partial}
\newcommand{\nno}{\nonumber}
\newcommand{\bse}{\begin{subequations}}
\newcommand{\ese}{\end{subequations}}

\begin{document}
\title{Pre-heating in the framework of massive gravity}
\author{Debaprasad Maity \footnote{debu.imsc@gmail.com}}
\affiliation{Department of Physics, 
Indian Institute of Technology, Guwahati, India}

\begin{abstract}
In this paper we propose a mechanism of natural preheating of our universe by introducing an 
inflaton field dependent mass term for the gravitational wave for a specific class of massive gravity theory. 
For any single field inflationary model, the inflaton must go through the oscillatory phase after the end of inflation. 
As has recently been pointed out, if the gravitational fluctuation has inflaton dependent mass term,
there will be a resonant amplification of the amplitude of the gravitational wave during the oscillatory phase of inflaton. 
Because of this large enhancement of the amplitude of the gravitational wave due to parametric resonance, 
we show that universe can naturally go through the pre-reheated phase with minimally coupled matter field. 
Therefore, during the reheating phase, there is no need to introduce any arbitrary coupling between the matter
field and the inflaton .

\end{abstract}

\maketitle

\newpage
\section{Introduction}\label{intro}
During the last decade, observational cosmology \cite{PLANCK} has reached to a level, through which
we can go backward in time, and see our universe 
with a great precession almost near the beginning of our universe. By now, we have understood very well the mechanism
of how the observed very tiny fluctuation ($\delta T/T \simeq 10^{-5}$) in temperature
in the Cosmic Micro-wave Background (CMB) sky plays the roll of seed field for the structure of the current universe. 
So far inflation, which is believed to have happened within a very short period of time after the big-bang, 
is believed to be the best theoretical explanation 
of the origin of those tiny fluctuation. The inflation was first proposed  
 \cite{guth,linde1,steinhardt} to explain some important conceptual problems persisted in the standard hot
big-bang cosmology. In spite of its simplicity, it was soon realised that constructing such 
a model within the known theory is very difficult. Main problems in constructing an inflation model
are the mechanism of how it has happened, the end of inflation, how the universe gets reheated after such a super cooling phenomena,
and most importantly the origin of inflaton itself. In addition to all those, there
generally exist many issues within a particular model under consideration. 
Living with many fundamental questions, in this paper we
will ask the question as to how reheating happened after the end of inflation. Immediately after
the proposal of inflation, this question had naturally emerged and studied in detail.  
There exist lot of studies on this particular mechanism \cite{branden,kofman}, which we generally call pre-heating.
In this paper we will introduce a new mechanism in the framework of a certain class of gravity theories where 
the gravitational fluctuation acquire inflaton field dependent effective mass term either from the original theory such as massive
gravity theory or from some higher derivative gravity theory. In this paper we will be focussing on 
a class of massive gravity theory proposed in \cite{dubovsky}.  
In the aforementioned class of theories our mechanism will be generic in the sense 
that for any inflationary model, after the end of inflation, 
the inflaton field generically oscillates, which in turn will provide the oscillating mass term for the graviton, and
resonant amplification in the matter sector will naturally happen without invoking any further
coupling parameters. In order to realise this, in this paper we will consider a class of 
ghost free massive gravity model \cite{dubovsky}, where we only have one extra degree 
of freedom in addition to the usual two tensor polarization modes. And we will identify the additional
degree of freedom to be the inflaton. We will see that in this class of models the gravitational 
fluctuation gets mass which is dependent on that extra scalar degree of freedom.  
The general physical mechanism to transfer the energy during pre-heating period  
is to couple the matter field as a probe with the background oscillating inflaton field.  
In the usual preheating scenario, one generally couples the matter fields with the inflaton field in an {\it ad hoc} manner, 
and try to tune those unknown coupling constant to have explosive particle production through parametric resonance.  
Here we will show that because of inflaton dependent mass term for the graviton fluctuation how the 
universal minimal coupling of the matter field
with the gravity is sufficient to reheat our universal after the end of inflation. 
The idea was first put forward in the reference \cite{CS-pre}, where the inflaton field is 
assumed to be coupled with the higher derivative gravitational Chern-Simons term. 
However, as we have explained in \cite{debu-pankaj}, it is very difficult to have successful preheating with 
the Chern-Simons coupling. It was shown to be crucially dependent upon the large initial amplitude of the gravitational fluctuation
during reheating. However, this large value of initial amplitude is found to be very difficult to achieve. Thanks to the reference \cite{misao}, 
which pointed out the possibility of generating large amplitude of the gravitational wave by parametric resonance
considering the inflaton field dependent mass term of the graviton. Purpose of the proposal was to create large scalar to tensor ratio 
without violating Lyth bound \cite{Lyth}. We are interested in preheating. Therefore, to achieve that we will 
have sequence of parametric resonance phenomena. We will see how the parametric resonance of the gravitational wave naturally 
leads to the parametric resonance in the matter sector though gravitational coupling. 
In this paper we will try to consider a minimal model \cite{dubovsky} to realise this mechanism.  

We structured our paper as follows: in the next section-II we will introduce a class of massive gravity theory,
which we will be studying. In the subsequent section-III, we will do the linear perturbation analysis 
in the gravity and the matter sector. As there in no direct coupling of the matter field with the inflaton,
we need to take into account the higher order term in matter field equation. In section-IV, 
we numerically solve those equations of motion and show that how the non-linear interaction term
between the gravitational fluctuation and the matter field is responsible for the explosive particle production.
The main driving force behind our mechanism is the successive resonance phenomena. The first resonance happens
in the gravitational sector because of the oscillatory mass term and the resonance in the matter field happens in the 
resonant gravitational background. Because of this resonance in a resonant background, the particle production turned out 
to be much more efficient in our model than in the usual re-heating scenario. In the final section,
we will make some final remarks on our mechanism and also discuss about some future works.

\section{Massive Gravity Model} 
Massive gravity theory has a long history starting from Fierz-Pauli \cite{pauli}, who first proposed 
a linearised theory of massive spin two field. Subsequently it has been shown that the linear theory
has some pathological behaviours which can not be cured in the linear regime. Therefore, non-linearity
has been introduced at the action level in order to have a consistent interacting spin two field theory \cite{gabadadge}.
One of the interesting facts of the non-linear modification proposed in \cite{gabadadge} is its near unique structure 
when the Lorentz invariance is imposed. But subsequently, the theory has been proved to be difficult 
to make phenomenologically viable because of its low energy cut off \cite{izumi}.  However, modificantion can be made 
by suitably modifying the Vainshtein scale to extend its regime of validity \cite{massive}.
Constructing model would be easer if one relaxes the Lorenz invariance. Therefore, further modification
has been proposed to violate Lorentz invariance in a controlled manner such that the theory remains well defined and also becomes 
cosmologically viable \cite{dubovsky,lin,pilo}.
As we have already mentioned in the introduction, in this paper we will be studying a particular
class of models proposed in \cite{dubovsky}. We want to demonstrate our mechanism considering the
minimal model where we have gravity with one inflaton field. In order to achieve this one considers
the following internal symmetries in the goldstone boson sector of gravity
\bea
\phi'^i = R^{i}_{j} \phi^{j}~~~~;~~~~\phi'^i= \phi^i + F^i(\phi^0),
\eea  
where, $R_{ij}$ is the rotation matrix in the field space, and $F^i(\phi^0)$ is arbitrary
function of $\phi^0$.
With this assumption we can write most general action for massive gravity as follows
\bea 
{\cal L} =  M_p^2  R + {\cal L}(X,Y^{ij},\Box\phi^0,\cdots) ,
\eea
where, ${\cal L}(X,Y^{ij})$ is the most general action for the goldstone boson
fields $(\phi^0,\phi^i)$ in accord with the aforementioned internal symmetries. This potential term will contribute
to the mass term for the graviton in the unitary gauge.
Based on the aforementioned residual symmetries in the goldstone sector, one can write the following scalar
quantities which we introduced above in ${\cal L}(X,Y^{ij})$,
\bea
X = g^{\mu\nu} \pr_{\mu}\phi^0 \pr_{\nu}\phi^0 ~~~;~~~Y^{ij} = g^{\mu\nu} \pr_{\mu}\phi^i \pr_{\nu}\phi^j - 
\frac{(\pr_{\mu}\phi^0 \pr^{\mu}\phi^i )(\pr_{\alpha}\phi^0 \pr^{\alpha} \phi^j)}{X} .
\eea    
Since the theory has the symmetry $\phi'^i= \phi^i + F^i(\phi^0)$, by using that
one can remove three propagating degrees of freedom from the $\phi^i$ with its
fixed vacuum expectation value $\langle\phi^i\rangle = x^i$. By using the Hamiltonian analysis also
it has been shown that the above theory has only three degrees of freedom. We identify the
extra degree of freedom $\phi^0 = \phi$ as an inflaton in our subsequent analysis. 
For the sake of our numerical computation we will consider the model of 
inflation from our previous paper \cite{debu-pankaj}. In compatible with the above
symmetries we consider the following Lagrangian,
\bea
 {\cal L}_{lv}(X,Y^{ij},\Box\phi,\cdots) ~=~ \frac 1 2 X - \frac 1 {2 s^3} M(\phi) X \Box \phi
- V(\phi) - \frac 8 9 M_p^2 m(\phi)^2
 \frac{{\bar Y}^{ij}{\bar Y}_{ij}}{Y^2},
\label{action}
\eea
where, we define 
\bea
V(\phi) = \Lambda^4 \left(1 -\cos \left ( \frac {\phi}{f}\right)\right)~~~;~~~
{\bar Y}^{ij} = Y^{ij} - 3 \frac {Y^{ik}Y_{k}^{i}}{Y^2}  ~~~;~~~ Y = \delta_{ij} Y^{ij}.
\eea 
To the leading order, the crucial difference between the above Lagrangian and the usual massive gravity 
theory in the literature is the extra scalar field dependent mass term \cite{misao}. 
We are not going to study the theoretical aspects of this kind if mass term in this paper.
As one can easily identify, this is the term which plays a crucial role for our mechanism to work. 
According to our previous discussion, we will study phenomenological aspects of this term in the cosmological context. 
If one considers the action in unitary gauge, the last term in the above action will play
like the mass term of the graviton. $s$ is the dimensionful parameter.
$M(\phi)$ is the some specific dimensionless function of the inflaton field.
$f$ is the decay constant. $\Lambda$ can be identified with the scale of inflation.
It is also interesting point to note that the mass term will not play the
role in the background dynamics. 

 We would like to emphasize that in this work we are only interested in the 
regime where the inflaton field coherently oscillates after the end of inflation.
To have successful reheating this is one of the important requirements for 
any model of inflation. Therefore, our final conclusion of this work is valid
for any model of inflation. The new ingredient that we have in the Lagrangian is
the extra term related to the mass of the graviton. The 
extra mass term is constructed in such a way that it only contributes 
to the mass of the gravitational fluctuation. 
Similar situation happens for gravitational Chern-Simons term coupled with the inflaton. 
The crucial difference is that the Chern-Simon term provides oscillating anti-dissipation term 
for the dynamics of the graviton. On the other hand here we have the oscillating mass term. 
Thanks to the reference \cite{misao}, in which the authors have already shown that
in the oscillating inflaton background the amplitude of the graviton fluctuation undergoes
resonant amplification. In this paper will show how that resonant amplification helps
to have resonant amplification of the matter field through minimal gravitational coupling.
In the subsequent analysis we will show how this mechanism works.

\section{Tensor Perturbation and Pre-heating}

 In this section we will analyse the tensor and matter perturbation in the oscillating
inflaton background. As is well known, 
in the usual pre-heating scenario, in order to create the matter field  from the 
inflaton, one invokes different types of arbitrary coupling parameters in the matter Lagrangian,
\bea
{\cal L}_{mat} = \frac 1 2  \pr_{\mu}\theta \pr^{\mu} \theta - m_{\theta}^2 \theta^2 
-\frac 1 2 g^2 \phi^2 \theta^2 + \frac 1 2 \chi R \theta^2 + \cdots
\eea
where, $\chi$ and $g$ are the coupling parameters. In the usual Einstein gravity, since
there is no parametric resonance in the gravity sector after inflation, one needs to introduce
above coupling parameters for every matter field to obtain the oscillating mass term 
for the matter field under consideration. Parametric resonance will occur because
of those coupling parameters. This in turn will provide us the explosive particle production.
What is novel of our mechanism is that, for the purpose of explosive particle production, we
do not need to introduce any arbitrary coupling constants for every matter fields.
We will show in our following analysis, how minimal gravitation coupling is sufficient
for this purpose. The only parameter that we need in our study is the gravitational mass
$m(\phi)$, which we have introduced before in the massive gravity action. Therefore, for 
our purpose, we set  
\bea
g =0 ~~;~~\chi=0
\eea  
As we have already mentioned, at the linear order in perturbation,
the extra mass term of the gravitational action will only contribute to the mass of the gravitational fluctuation.
The background evolution is governed by the inflaton field $\phi$ with the following metric
\bea
ds^2 = -dt^2 + a(t)^2(dx^2 + dy^2 +dz^2).
\eea  
After the end of inflation, the inflaton start to oscillate which is known to be responsible for the
reheating of our universe. For current study we will consider the 
\begin{figure}
\includegraphics[width=2.00in,height=1.50in]{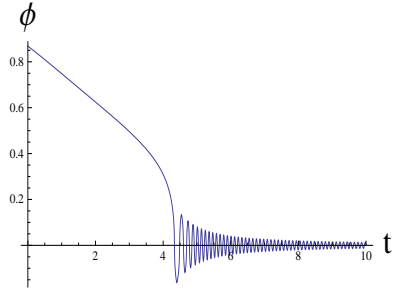}
\caption{\label{fig1} Background evolution of the inflaton field $\phi$ in cosmological time t. 
We measure the value of $\phi$ in unit of $f$ and cosmic time $t$ in unit of $s$.}
\end{figure}
solution of the oscillating inflaton from our reference \cite{debu-pankaj} as shown in the fig.\ref{fig1}. As we
have emphasized before, our mechanism does not depend on the particular model of inflation under consideration. As long as we have 
oscillating background inflaton, we will have broad parametric resonance for the gravitational fluctuation for a
wide range of momentum modes depending upon the oscillation frequency of the inflaton and the coupling paramter $\lambda$.
At the linear order, 
the scalar field fluctuation will follow the same dynamics as before. Since our mechanism
of pre-heating is mainly dependent on the dynamics of gravitational fluctuation, we consider
tensor perturbations as follows
\bea
ds^2 =  -dt^2 + a(t)^2(\delta_{ij} + h_{ij}) dx^i dx^j
\eea
Apart from inflaton we only have two tensor polarization degrees of freedom. Therefore, 
we can always choose the following transverse and traceless gauge condition for the tensor perturbation,
\bea
\partial_i h^{ij} = 0~~;~~\delta_{ij} h^{ij} = 0 .
\eea
The fourier mode function of $h_{ij}$ defined as
\bea \label{gwmode}
h_{ij}(t,x) = \int \frac {d^3 {\bf k}}{(2 \pi)^{3/2}} \sum_{s=1,2} [e^s_{ij}({\bf k}) {\tilde h}^s ({\bf k},t) e^{ i {\bf k}\cdot{\bf x}} + h.c.],
\eea
Where, $e^s_{ij}({\bf k})$ are the two independent polarization tensor of the gravitational wave.
To the linear order in cosmological perturbation theory, the Fourier transformed gravitational wave equation 
turns out to be
\bea \label{gwave}
\ddot{{\tilde h}}^s + 3 H \dot{{\tilde h}}^s + \left(\frac {k^2}{a^2} + m(\phi)^2 \right) {\tilde h}^s =0 .
\eea
This equation is modified Mathieu equation with the oscillating inflaton background.
The usual re-heating phenomena is based on this same form of equation of a
matter field. Analytic study of this equation has been done quite extensively
in the context of pre-heating. The important difference here is that, resonant 
production of gravitational wave initiate the pre-heating process through the 
minimal coupling of the matter field.

By using the properties of the gravitational wave, the form of the scalar field equation becomes
\bea
{\ddot {\theta}} + 3 H \dot{\theta} -\frac {1}{a^2} \nabla^2 \theta + \frac {h^{ij}} {a^2} \partial_i\partial_j \theta + m_{\theta}^2  \theta = 0 .
\eea 
As one notices that in order to have gravitational effect we restrict up to the second order term.  
Considering the following Fourier transformation for the scalar field,
\bea
\theta(t,x) = \int \frac {d^3 {\bf k}}{(2 \pi)^{2/3}} [a_{\bf k} ~{\tilde \theta}({\bf k},t) e^{ i {\bf k}\cdot{\bf x}} + h.c.],
\eea
the mode equation for ${\tilde \theta} ({\bf k})$, take the following from 
\bea \label{thetaeq}
\ddot{\tilde \theta} + 3 H \dot{\tilde \theta} + \left(\frac {k^2}{a^2} + m_{\theta}^2 \right) {\tilde \theta} = 
 - \left(\frac {2}{\pi}\right)^{3/2} \frac {1}{a^2} 
\int \sum_{s} k'_{i} k'_{j}~e^s_{ij}({\bf k}-{\bf k}') {\tilde h}^{s}({\bf k}-{\bf k}')~ {\tilde \theta}({\bf k}')~ d^3 {\bf k}' .
\eea 
In the integral part of the above evolution equation for the scalar mode, we consider the evolution of a particular
scalar mode of ${\bf k}$. For a fixed vector ${\bf k}$, one can parametrize the two tensor polarization $e^s_{ij}({\bf q})$ modes
of the gravitational wave in the above integral as,
\bea
e^{ij}_1 ({\bf q}) &=& \epsilon^{i}_1({\bf q}) \epsilon^{j}_2({\bf q})+\epsilon^{i}_2({\bf q})\epsilon^{j}_1({\bf q}), \nno \\
e^{ij}_2 ({\bf q}) &=& \epsilon^{i}_1({\bf q}) \epsilon^{j}_1({\bf q})- \epsilon^{i}_2({\bf q})\epsilon^{j}_2({\bf q}). \nno
\eea
where, in a particular basis we write
\bea
{\mbox{\boldmath$\epsilon$}}_1({\bf q}) &=& \left\{-\frac {{\bf q}_{y}}{q_{xy}}, \frac {{\bf q}_{x}}{q_{xy}},~ 0 \right\}, \nno \\
{\mbox{\boldmath$\epsilon$}}_2({\bf q}) &=& \left\{\frac {{\bf q}_{x} {\bf q}_{z}}{q_{xy} q} ,~ \frac {{\bf q}_{y} {\bf q}_{z}}{q_{xy} q},~- \frac {{\bf q}_{xy} }{q} \right\}, 
\eea
considering ${\bf q} = ({\bf k}' - {\bf k})$. As we have mentioned before, for a fixed scalar mode 
$\theta_{\bf k}$ with momentum $ {\bf k} = k \hat{{\bf z}}$,
if we parametrize the ${\bf q} = \{ k' \sin \alpha \cos \beta , k' \sin \alpha \sin \beta , k - k' \cos \alpha \}$. 
With the above parametrization, the equation of motion for the scalar mode function turned out to be
\bea \label{thetaeqmod}
\ddot{\tilde \theta} + 3 H \dot{\tilde \theta} + \left(\frac {k^2}{a^2} + m_{\theta}^2 \right) {\tilde \theta} = 
 2 \pi \left(\frac {2}{\pi}\right)^{3/2} \frac {k^2}{a^2} 
\int \frac{k'^4 \sin^3 \alpha ~  {\tilde h}({\bf k}-{\bf k}')} {k^2 + k'^2 - 2 k k'  \cos \alpha}~ {\tilde \theta}({\bf k}')~ d\alpha dk' .
\eea
Now this is a modified Mathieu equation, where the oscillatory background is set by all possible modes of the 
gravitational wave $ {\tilde h}({\bf k}-{\bf k}')$ , which satisfy the following equation  , 
\bea
\ddot{{\tilde h}}^s + 3 H \dot{{\tilde h}}^s + \left(\frac {k^2 + k'^2 - 2 k k'  \cos \alpha}{a^2} + m(\phi)^2 \right) {\tilde h}^s =0 .
\eea 
In order to understand the behaviour of these equations, we
will consider two different limit on the possible value of the scalar mode momentum. First let us
consider large $k$ limit compared to $k'$. The modes in this range will contribute the most.
In this limit, the solution of ${\tilde h}({\bf k}-{\bf k}')$ becomes
approximately independent of $k'$. Therefore, the above equation for the scalar mode function will take the following
simplified form after performing the angular integral, 
\bea
\ddot{\tilde \theta} + 3 H \dot{\tilde \theta} + \left(\frac {k^2}{a^2} + m_{\theta}^2 \right) {\tilde \theta} = 
 \frac {8 \pi}{3} \left(\frac {2}{\pi}\right)^{3/2} \frac { {\tilde h}({\bf k})}{a^2} 
\int k'^4 ~ {\tilde \theta}({\bf k}')~  dk' .
\eea 
To understand the scale around which we will have the resonant amplification within the perturbative limit,
let us introduce the natural time scale $s^{-1}$ coming from the period of oscillation of the inflaton field
after the end of inflation as mentioned in fig.\ref{fig1}. If we measure our time in term of $s$ unit, 
then the above scalar mode equation turns out to be
\bea \label{kovers}
\ddot{\tilde \theta} + 3 H \dot{\tilde \theta} + \left(\frac {k^2}{s^2 a^2} + \frac {m_{\theta}^2}{s^2} \right) {\tilde \theta} = 
 \frac {8 \pi}{3} \left(\frac {2}{\pi}\right)^{3/2} \frac {s^3 {\tilde h}({\bf k})}{a^2} 
\int \frac {k'^4}{s^4} ~ {\tilde \theta}({\bf k}')~ \frac {dk'}{s} .
\eea 
For the purpose of our analytic discussion, we set the scale factor $a(t) = 1$. This is a reasonable assumption during 
pre-heating, as the scale factor remains almost constant. We also use this fact 
for our numerical analysis in the later section. At this point we also would like
to point out that the combination $s^3 {\tilde h}({\bf k})$ is dimensionless as can bee seen from the fourier expansion
of the gravitational wave eq.(\ref{gwmode}). Therefore, our perturbative analysis will be under control as long as 
one satisfies $s^3 {\tilde h}({\bf k}) < 1$. From the usual properties of Mathieu equation, the broad parametric
resonance will occur if the coefficient of ${\tilde \theta}({\bf k}')$ in the integral part of the above scalar mode equation is larger than unity.
Therefore, from the scaling properties of momentum $k'$ and ${\tilde h}({\bf k})$, we can set the following qualitative condition 
between the magnitude of the gravitational field and the momentum modes contributing to the scalar particle production as follows,
\bea
s^3 {\tilde h}({\bf k}) \left(\frac {k'}{s}\right)^5 \gg 1 \implies   \left(\frac {s}{k'}\right)^5 \ll (s^3 {\tilde h}({\bf k}) 
\eea
Furthermore, within the perturbative regime, the modes which will contribute to the broad parametric resonance for the 
scalar amplitude should satisfy the following condition,
\bea \label{kk1}
\left(\frac {s}{k'}\right)^5 \ll s^3 {\tilde h}({\bf k}) <1 .
\eea
However, if we consider, $k'$ to be larger than $k$, following the same procedure we 
have discussed so far, the equation of motion for the scalar mode function will turn out to be 
\bea \label{kovers2}
\ddot{\tilde \theta} + 3 H \dot{\tilde \theta} + \left(\frac {k^2}{s^2 a^2} + \frac {m_{\theta}^2}{s^2} \right) {\tilde \theta} = 
 \frac {8 \pi}{3} \left(\frac {2}{\pi}\right)^{3/2} \frac {k^2}{s^2 a^2} 
\int \frac {k'^2}{s^2} ~s^3 {\tilde h}({\bf k}') {\tilde \theta}({\bf k}')~ \frac {dk'}{s} .
\eea 
Following the similar argument as we have discussed so far, the constraint on the momentum modes will be as follows
\bea \label{kk2}
\left(\frac {s}{k}\right)^2 \left(\frac {s}{k'}\right)^3  \ll s^3 {\tilde h}({\bf k}') < 1, ~~~\mbox{for}  ~~~ k \ll k' . 
\eea
At this point would like to emphasized the fact that the conditions for the momentum modes
eqs.(\ref{kk1}) and eq.(\ref{kk2}) for two different limit of the scalar momentum $k>>k'$ and $k<<k'$ respectively, 
are within the perturbative limit. 
The usual parametric resonance will lead the gravitational amplitude to be in the non-perturbative regime. 
In our present analysis we will confine ourselves in the perturbative regime by limiting the amplitude of the
gravitational wave to be less than unity. However, non-linear analysis must be needed to understand complete dynamics
of the system. Furthermore, it may also be possible that back-reaction of the scalar particle production into the gravitational
wave dynamics keep it always in the perturbative regime. We defer all these issues for our future studies. 

We have not introduced any direct coupling of the matter field with the inflaton. Therefore, it important
to note that in the dynamics of the matter field $\theta$, the gravitational fluctuation will play
the role at the first non-linear order. However, in order to realise our mechanism,
not only we need to choose suitable mass term $m(\phi)$ for the graviton fluctuation, but also
we need to be careful about its effect during the inflation. In the usual massive gravity theory, 
modification is tractable in the infra-red limit
as the theory remains perturbative in nature. Non-linearity of the theory depends crucially upon the mass term of the graviton. 
In the current construction, we consider gravitational mass is dependent on the inflaton field itself. 
Therefore, one needs to study the strong coupling aspects of the theory which we will defer for our future study.
Motivated by the reference \cite{misao} we assume, in order not to have 
large power suppression of the gravitational wave during inflation, 
\bea
m(\phi)^2 = \frac {\lambda \phi^2}{1+ \left(\frac {\phi}{\phi*}\right)^4}  ,
\eea
where $\phi*$ is the value of $\phi$ at the end of inflation. Such a functional form will 
lead to very small mass of the gravitational wave during inflation, since inflaton field
will have very large constant vacuum expectation value during inflation. One can easily compute
the tensor power spectrum during inflation with almost constant mass limit of the graviton to be
\bea
P_{h} = \frac {2 H^2} {\pi^2 M_p^2} \left(\frac {k} {a H}\right)^\frac {2 m(\phi)^2}{3 H^2} .
\eea
Therefore, as one can see that due to the specific choice of our mass funtion $m(\phi)$, modified part of
the tensor power spectrum due to the mass term is highly suppressed. However, after the end of inflation, since the
inflaton undergoes damped oscillation around the minimum of it's effective potential, 
the amplitude of the inflaton will be very small compared to $\phi*$. Therefore, during oscillating phase, 
we can approximate,
\bea
m(\phi)^2 = \lambda \phi^2,
\eea
where $\lambda$ is dimensionless coupling constant. observing the above expression for the inflaton 
dependent mass term, one would expect to get back the usual massless gravity theory long after the end of inflation. 
For a class of massive gravity theory \cite{dubovsky}under consideration, it has been shown that
it reproduces usual gravity theory in the massless graviton limit. In this paper we will be 
restricted to a particular class of minimal massive gravity models where the number of degrees of freedom is three.
\begin{center}
\begin{figure}
\includegraphics[width=6.00in,height=1.50in]{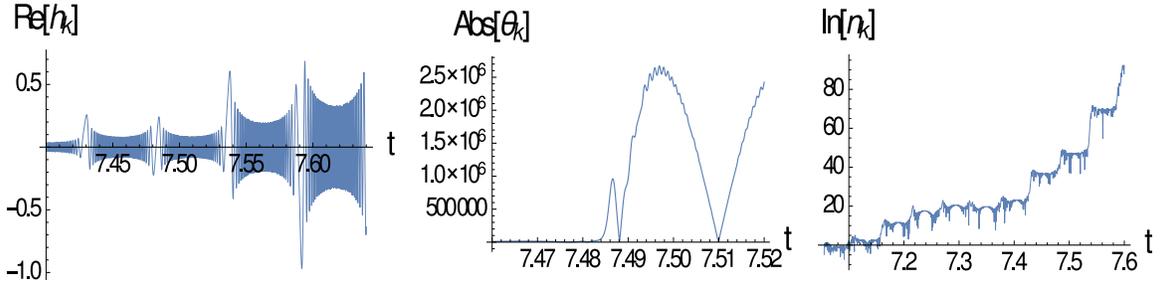}
\caption{\label{fig2} Behaviour of gravitational wave $Re[h_k]$ in unit of $s^3$, the matter field $\theta_k$ and particle number 
density $n_k$ in time t which is again measured in unit of s. We have plotted for $k_x = 0, k_y = 170, k_z=130$ in unit of s. 
We chosen the dimensionless graviton mass parameter $\lambda =1$. We also checked that as we decrease the initial amplitude of the
gravitational wave, the resonant production of the gravitational wave happens at later time which leads to the
particle production at late time.}
\end{figure} 
\end{center}
Now considering the gravitational wave equation with the above form of the gravitational mass term
\bea \label{gwave}
\ddot{{\tilde h}}^s + 3 H \dot{{\tilde h}}^s + \left(\frac {k^2}{s^2 a^2} + \frac {\lambda \phi^2}{s^2} \right) {\tilde h}^s =0 .
\eea
As usual, we consider the time scale in unit of $s$.
As has been extensively discussed in \cite{kofman}, and we also have observed from our numerical 
analysis the successive enhancement of the amplitude of the gravitational wave 
happens at the time of zero crossing of the inflaton field, and it oscillates with almost 
the same amplitude in between the two successive crossings. Therefore, analytic treatment of this 
equation suggested in \cite{kofman}, can be done by the method of successive 
parabolic scattering. The stochastic resonance certainly will lead the perturbative gravitational
amplitude into the non-perturbative regime. Therefore, one needs to go the non-linear regime.
However, in the present paper we will restrict our study in the 
perturbative regime such that the approximate energy density of the gravitational wave \cite{kofman}
\bea
\rho_{h} \simeq  \frac{\lambda |\phi|}{ \pi^2 a^3} \int_{0}^{\infty} ~dk ~k^2 ~|\beta_k|^2, 
\eea
where, $\beta_k$ is the amplitude of the negative frequency part of the full solution, is smaller than the inflaton energy density. 
Because of resonance, the approximately expression for $|\beta_k|^2 \simeq \frac {1}{2} e^{2 \mu_k s t}$, where the exponent $\mu_k$,
defined for a particular instant of zero crossing of the inflaton field is given as
\bea
\mu_k(t) = \frac {1}{2 \pi} ln\left( 1 + 2 e^{- \kappa^2} - 2 \sin  \theta e^{- \frac{\kappa^2}{2}} \sqrt{1+e^{- \kappa^2}}\right).
\eea 
The expression for $\kappa^2 = \frac {\pi k^2}{a^2 \lambda s \phi(t)}$, and $\theta$ is random phase within $(0, 2 \pi)$.
Therefore, on can see the exponential enhancement of the gravtitional field amplitude. In order to be 
in the perturbative regime, we have to consider $t s$ begin very small. 

One may notice, in our current analysis the eq.(\ref{gwave}),in the dynamics of the
gravity wave, we ignored the back-reaction of the matter field. It can be easily understood that
at least at the initial stage of the pre-heating process, the back-reaction should be small. But it must
be very important after sufficient number of matter particle produced. The full
analysis taking into account the back-reaction and rescattering of the particles will be considered in our subsequent publication.
However, as we pointed out, the matter particle production will happen at the non-linear level. Thanks to
the parametric resonance that has happened in the gravitational sector. Because of this resonance, we will show,
how efficient the particle production is in our mechanism compared with the usual reheating mechanism. In order
to quantify the amount particles that has been produced for $\theta_k$ with momentum ${\bf k}$    
we use the following well known expression of particle number density \cite{kofman} 
\bea
n_k = \frac {\omega_k}{2} \left( \frac {|\dot{\tilde{\theta}}({\bf k})|^2} {\omega_k^2} + {|{\tilde{\theta}}({\bf k})|^2}\right) - \frac 1 2 .
\eea
Where, $\omega_k$ is the time dependent frequency of a particular mode ${\bf k}$. 
\bea
\omega_k =  \sqrt{\frac {k^2}{a^2} + m_{\theta}^2} = \sqrt{{k_{phy}^2} + m_{\theta}^2},
\eea
where $k_{phy}$ is the physical momentum. In the above derivation of the number density, we have used the 
following canonical quantization condition for the scalar field mode,
\bea
{\dot {\tilde{\theta}}^*}(t) {\tilde{\theta}}(t) - {\tilde{\theta}}^*(t){\dot {\tilde{\theta}}}(t) = i
\eea

In the next section will study numerically two equations of motion and try to compute the particle number
density for a particular mode. For simplicity of our numerical computation,
we fix the value of $k_x, k_y$ and do the numerical integration for the mode equation of $\theta_k$
along the $k_z$. 

\subsection{Numerical analysis and particle production}

As we have already discussed before, we are only interested in the oscillatory regime of the inflaton, our final 
conclusion does not really depend on a specific model of inflation. 
In order to get the background of oscillatory inflaton for our 
numerical calculation, we consider our specific
model of modified natural inflation \cite{debu-pankaj}. We choose 
$M(\phi) ={\sin^7 x} \left[1-\cos x \sin^2 x \right]^{14}$, where $x = \phi/f$.
For this specific form of the function, we had the number of e-folding ${\cal N} = 50$,
and the value of the spectral index is $n_s=0.96$. In order to achieve the aforementioned cosmological
quantities compatible with the observations, we found the values of the other
parameters to be $f = 0.84 M_p, s = 2.25 \times 10^{-6} M_p, \Lambda = 0.011 M_p $. 
For our numerical purpose we set $t=0$ as the beginning of the inflation. After 
the end of inflation at around $ t \sim 4 $ cosmic time in unit of $s$, the 
inflaton field will start to have coherent damped oscillation as shown in the fig-\ref{fig1}.
At this point, we are again emphasizing the fact that our analysis and
the final result is confined within the oscillatory regime of the inflaton field. 
The oscillation after inflation is generic to any inflationary model. Therefore, our qualitative results
are insensitive to the specific model of inflation under consideration. 
For the sake of our numerical analysis, at time $t = 6$ we normalize the cosmological 
scale factor $a(6) = 1$. For a wide range of initial amplitude
of the gravitational wave, we found the resonant production of the gravitational wave.
More over this resonant production of gravitational wave will trigger the pre-heating 
phase of the universe through the matter fields with the universal gravitational coupling. 
In the inflationary background, to quantize a field we choose the usual
Bunch-Davis vacuum, 
\bea
{\tilde \theta}(k) = \lim_{t \rightarrow 0 }  \frac 1 {\sqrt{2 k}} e^{- i k \int \frac 1 {a(t)} dt} \left(1 + \frac i { k \int \frac 1 {a(t)} dt}\right).
\eea
In the above expression, we also assume approximately the de-Sitter vacuum. As we see, lower the value of a particular mode $k$, 
larger the initial amplitude in the quantum vacuum, leads to the earlier production of matter particle of 
that particular modes after the end of inflation. 
As a particular mode starts evolving from its quantum vacuum defined in the far past and deep inside the Hubble volume during inflation,
its amplitude will start to decrease due to Hubble friction and finally gets frozen after the horizon exit. 
Therefore, during reheating the amplitude at the horizon exit will set the initial condition for a 
particular mode of the gravitational wave to evolve.
From the aforementioned arguments, we see that the parametric resonance for the matter field will happen 
mostly for the long wavelength modes compared to the Hubble length. We can clearly see from the fig-\ref{fig2} that because of the
resonant gravitational wave production triggered by its oscillatory mass term,
minimal coupling of the matter field with the gravity is sufficient to reheat the universe. 
\begin{center}
\begin{figure}
\includegraphics[width=5.00in,height=2.00in]{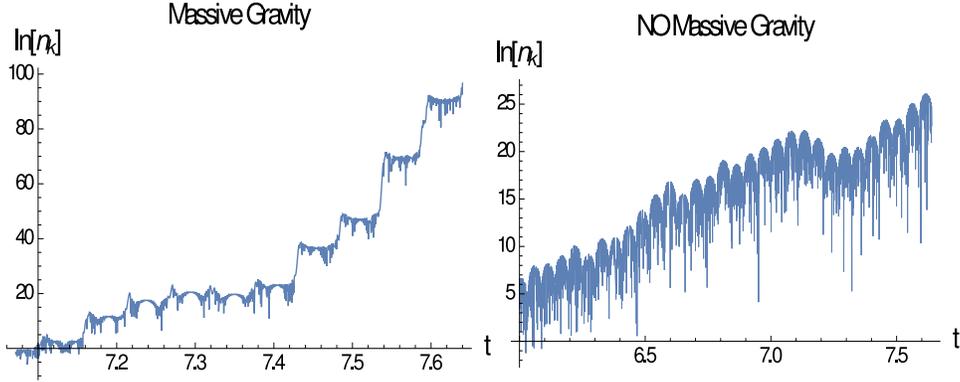}
\caption{\label{fig3} Comparing the particle production for matter field $\tilde{\theta}(k)$ in massive gravity scenario (left panel) 
with the usual pre-heating scenario(right panel) considering ${\cal L}_{int} \sim~ g~ \phi^2 ~\theta^2$ coupling term. Plot for the particle
density $n_k$ in time t which is measured in unit of s. We take $k_x = 0, k_y = 170, k_z=130$ in unit of s. 
Choose $\lambda = g = 1, m_{\theta} = 0.1 s $. }
\end{figure} 
\end{center}
we also want to compare our result with respect to the usual pre-heating scenario \cite{kofman}.
At the linear order in perturbation theory, in our current scenario we want to see how efficient the particle production is 
in comparison with the usual pre-heating scenario \cite{kofman}. One can clearly see the difference between the 
distinct effect of parametric resonance on the gravitational fluctuation $\tilde{h}(k)$ and the matter field 
fluctuation $\tilde{\theta}(k)$ from the fig.\ref{fig2}. 
Important point to mention that in the scalar field sector, the parametric resonance happens in the 
resonant gravitational field background. Therefore, one can clearly see from the left panel of Fig.\ref{fig3} that the produced 
density of particle in the massive gravity framework is behaving like double exponential function in time.
The initial condition, we have used in our analysis is that there in no particle at $t= 6$ cosmic time in unit of $s$ after the inflation.
After the end of cosmic time $t=7.64$, so that the amplitude of the gravitational wave is in the 
perturbative limit, one can see from the fig.\ref{fig3} that the number of particle 
produced per unit volume in the massive gravity scenario is $\sim e^{75}$ times larger than that of the usual
pre-heating scenario for a particular mode. Therefore, in the massive gravity scenario, the particle production 
is much more efficient compared to the usual pre-heating scenario. 
It is possible that most of the energy of the inflaton field may easily be transferred 
through successive parametric resonance first to the gravitational sector and then to the matter sector within a very short period of time, and 
the usual reheating follows. The whole mechanism of particle production that we have discussed so far
is highly non-equilibrium in nature. Therefore, the elementary theory of reheating is still necessary to 
thermalization all the produced particle, and the universe to enter into the usual radiation phase.
Now, because of highly efficient particle production compared to the usual pre-heating scenario, 
if most of the inflaton energy is transferred to the relativistic matter particle, the usual 
standard model couplings among the produced particles may be sufficient to thermilize them. 
Therefore, we may not need to introduce any arbitrary coupling at all to have the successful reheating
of our universe after the inflation. Of course our current analysis is in the linear regime, and we have not
taken into the back-reaction and the re-scattering of the fields. We differ the detail studies
of all the aforementioned points for our future work. 

\section{Conclusion}
In this paper, we have tried to constructed a phenomenological model of pre-heating by introducing 
inflaton field dependent mass term for a certain class of massive gravity theory.
In that particular class of theories, we have three degrees of freedom because of some
special internal symmetries. For the purpose of our study, we identified the extra degree
of freedom to be the inflaton field.
During inflation, gravitational mass term will suppress the power in 
the gravitational wave spectrum. In order to avoid large suppression, we parametrize the mass 
function in such a way that during inflation, its value is very small. On the other hand, after the end
of inflation, the amplitude of the gravitational wave will be amplified
due to parametric resonance because of the mass term\cite{misao}. Now the question naturally arises 
is why have not we observed this large amplitude gravitational wave yet? 
Natural answer would be that this large amplitude gravitational wave will contribute
largely to reheat our universe after the inflation. In this paper we have shown that how this resonant gravitational
wave background can naturally lead to the pre-heating phase of our universe. 
Interestingly, to pre-heat the universe we do not need to introduce any arbitrary couplings between
the visible sector matter field and the inflaton field in an {\it ad hoc} manner.
The natural universal gravitation coupling is sufficient to reheat the universe.
At the linear order in perturbation theory, we also compared the efficiency of the particle production  
between our mechanism and the usual reheating mechanism \cite{kofman}. We found that because
of the resonant gravitational background, in our mechanism the particle production is highly efficient
compared to the usual scenario. This phenomena of resonance in a resonant background has not been
studied before at least in the cosmological context. It would very much interesting to study
this mechanism in detail in various physical situations.

Using the same mechanism that we discussed so far, we can study the production of chiral gravitational wave 
during preheating if we further introduce the Chern-Simons (CS) higher derivative
term \cite{CS,CS1} coupled with the inflaton field in the action, i.e.
\bea
{\cal L}_{CS} \simeq \frac {g} {f} \phi R {\tilde R} .
\eea
The inflaton field is now like a pseudo scalar axion, where 
\bea
R {\tilde R} &=&  \tilde{R}^{\nu}{}_{\nu}{}^{\alpha \beta} \, R^{\nu}{}_{\mu \alpha \beta}\, ~~;~~
\tilde{R}^{\nu}{}_{\nu}{}^{\alpha \beta} := \frac{1}{2}\, \epsilon^{\alpha \beta \gamma \delta} R^{\mu}{}_{\nu\gamma \delta}. \nonumber 
\eea
$\epsilon^{\mu\nu\alpha\beta}$ is the usual 4-dimensional Levi-Civita tensor. 
$g$ is the free coupling parameter. Since in the background of inflaton this
term violates parity, the left and right handed gravitational wave propagate 
differently according to the following equation 
\bea \label{gwave}
\left(M_p^2 + \frac {k ~ g ~\eta_A} {f ~a} \dot{\phi} \right)\left(\ddot{h}^k_A \right. &+& \left. 3 H \dot{h}^k_A + \frac {k^2}{a^2} h^k_A \right) + m(\phi)^2  h^k_A \nno \\ 
&=&- \frac {k~ g ~\eta_A}{f~ a} (\ddot{\phi} - H\dot{\phi})~\dot{h}^k_A ,
\eea
where $A = L,R$ are left and right circular polarization mode of the gravitational wave. 
Therefore, we will have resonant chiral gravity wave production. This effect may be very important to 
produce the leptogenesis in the early universe. As is well know \cite{witten}, if we couple gravity with the leptons,
the total lepton number current will no longer be conserved because of gravitational anomaly,
\bea
\partial_{\mu} j^{\mu} = \frac 3 {16 \pi^2} < R {\tilde R} > ,
\eea
where, $< R {\tilde R} >$ is the quantum expectation value of the Chern-Simons term. Therefore,
non-zero expectation value of this Chern-Simons operator will produce the leptogenesis. 
This mechanism has already been discussed in \cite{CS}.
However, in our model the resonant production of chiral gravitational wave will happen during oscillation 
period of the inflaton. Therefore, the maximum lepton number violating contribution will be coming during the preheating period. 
Currently we are looking into this issue in greater detail. We are also looking into the possibility this new reheating mechanism 
in a Lorentz invariant higher derivative gravity theory coupled with inflaton field. We know that the higher derivative
term such as Gauss-Bonnet term coupled with the inflaton field contributes an effective inflaton field dependent
mass term in the gravitational fluctuation. Currently we are also looking into that possibility.  

We have also pointed out that the specific class of massive gravity theories, we have considered for our study, need to 
be studied in detail. In massive gravity theory, one generally introduces the constant mass term for the graviton.
But here we have introduced extra field dependent mass term \cite{misao}. This particular
field dependent mass term should be studied with a special care. 
As one knows interacting massive spin two field coupled with other field such as our scalar field has several constraints to be satisfied. 
We keep all these questions for our future studies.

{\bf Acknowledgement}\\
We are very thankful to our HEP-group members to have vibrant academic discussions.

\end{document}